\documentclass[aps,prl,twocolumn,superscriptaddress,showpacs]{revtex4-2}
\usepackage{amsmath, amssymb,wasysym,graphicx}
\usepackage{appendix}
\usepackage[ruled]{algorithm2e}

\usepackage[utf8]{inputenc}
\usepackage[T1]{fontenc}
\usepackage{xcolor}

\usepackage{diagbox}
\usepackage{tabularx}

\IfFileExists{newtxtext.sty}
{\usepackage{newtxtext,newtxmath}}
{\IfFileExists{stix.sty}
	{\usepackage{stix}}
	{\IfFileExists{mathptmx.sty}
		{\usepackage{mathptmx}}{} } }

\usepackage{textcomp}

\usepackage{bm}

\usepackage{booktabs,siunitx}

\usepackage{color}
\definecolor{LinkColor}{rgb}{0.256,0.439,0.588}
\usepackage{hyperref}
\hypersetup{
	pdfauthor={good guys},
	pdftitle={good title},
	colorlinks=true,
	citecolor=LinkColor,
	linkcolor=LinkColor,
	urlcolor=LinkColor
}

\renewcommand{\vec}[1]{\mathbf{#1}}

\usepackage{pifont}
\usepackage[normalem]{ulem}

\newcommand{\etal}{\textit{et al.}}

\newcommand{\Tr}{\mathrm{Tr}}
\newcommand{\Det}{\mathrm{Det}}
\newcommand{\Diag}{\mathrm{Diag}}

\usepackage{multirow}

\begin{document}
\title{Network-Initialized Monte Carlo Based on Generative Neural Networks}
\author{Hongyu Lu}
\affiliation{Department of Physics and HKU-UCAS Joint Institute
of Theoretical and Computational Physics, The University of Hong Kong,
Pokfulam Road, Hong Kong SAR, China}

\author{Chuhao Li}
\affiliation{Beijing National Laboratory for Condensed Matter Physics and Institute
	of Physics, Chinese Academy of Sciences, Beijing 100190, China}
\affiliation{School of Physical Sciences, University of Chinese Academy of Sciences, Being 100190, China}

\author{Bin-Bin Chen}
\affiliation{Department of Physics and HKU-UCAS Joint Institute
of Theoretical and Computational Physics, The University of Hong Kong,
Pokfulam Road, Hong Kong SAR, China}

\author{Wei Li}
\email{w.li@itp.ac.cn}
\affiliation{Institute of Theoretical Physics, Chinese Academy of Sciences, Beijing 100190, China}
\affiliation{School of Physics, Beihang University, Beijing 100191, China}

\author{Yang Qi}
\email{qiyang@fudan.edu.cn}
\affiliation{State Key Laboratory of Surface Physics, Fudan University, Shanghai 200438, China}
\affiliation{Center for Field Theory and Particle Physics, Department of Physics, Fudan University, Shanghai 200433, China}

\author{Zi Yang Meng}
\email{zymeng@hku.hk}
\affiliation{Department of Physics and HKU-UCAS Joint Institute of Theoretical and Computational Physics, The University of Hong Kong, Pokfulam Road, Hong Kong SAR, China}

\begin{abstract}
We design generative neural networks that generate Monte Carlo configurations with complete absence of autocorrelation from which only short Markov chains are needed before making measurements for physical observables, irrespective of the system locating 
at the classical critical point, fermionic Mott insulator, Dirac semimetal, or
quantum critical point. We further propose a network-initialized Monte Carlo scheme 
based on such neural networks, which provides independent samplings and can
accelerate the Monte Carlo simulations by significantly reducing the thermalization 
process. We demonstrate the performance of our approach on the two-dimensional 
Ising and fermion Hubbard models, and expect it can systematically speed up the 
Monte Carlo simulations especially for the very challenging many-electron problems.

\end{abstract}
\date{\today}
\maketitle

{\it Introduction}\,---\,
Monte Carlo (MC) method is a widely used numerical method to investigate 
problems in statistical and quantum many-body physics. Although being generally polynomial when the sign problem is 
absent, the complexity of MC simulation can still be prohibitively 
high for certain challenging problems. In particular, 
quantum MC simulations of generic interacting fermion systems
\cite{BSS1981,Hirsch1983,Hirsch1985,Assaad2008,XuReview2019}, 
such as Hubbard model for correlated electrons \cite{Hirsch1983,Hirsch1985}, 
Holstein model for electron-phonon interaction~\cite{Scalettar1989,Noack1991,
ChuangChen2018,ChuangChen2019}, the spin-fermion model of non-Fermi liquid~\cite{XYXu2017PRX,ZHLiu2019PNAS,XuReview2019,WLJiang2021,
YuzhiLiu2021}, and the momentum space QMC for quantum moir\'e systems
\cite{XuZhang2021,JYLee2021,GPPan2021,XuZhang2021SC}, to name a few, 
have very high complexity that scales at least to $\sim \beta N^3$ where 
$\beta=1/T$ is the inverse temperature and the $N=L^d$ the total site number 
for a lattice model in $d$-dimension. Such type of MC approach for the 
many-electron problems is termed as determinant quantum Monte Carlo 
(DQMC) where the complexity comes from the matrix operation on the fermion 
determinant. The situation worsens when dealing
with critical points like the novel quantum critical points ubiquitously present 
in the metal-insulator transition, pseudogap and non-Fermi liquids, and dynamics 
and transport properties in high-temperature superconductivity, etc. This can largely 
be ascribed to that the widely-used Markov-chain Monte Carlo (MCMC) method 
with local updates (or the Metropolis algorithm) suffers from very long Markov-chain 
autocorrelation that scales with the system size to a high power due to the critical slowing 
down.


Although there exist powerful and highly efficient cluster/loop
update schemes for classical problems such as the Ising model~\cite{SwendsenWang1987,Wolff1989} and quantum
spin/boson systems such as Heisenberg and Bose-Hubbard models~\cite{Sandvik1999,Prokofev1998JETP,Prokofev1998PLA,
Sandvik2010}, these ingeniously designed algorithms do not exist for 
generic quantum many-body problems, especially for the interacting 
fermion systems, for example in DQMC. Therefore, the efforts for more 
efficient algorithms that could overcome these difficulties and boost the 
(quantum) MC simulations for the aforementioned systems are called for.

In recent years, the application of machine learning technique has achieved 
notable success in the studies of  systems ranging from 
statistical models to condensed matter and quantum materials research
\cite{Carleo2019,Carrasquilla2020,Bedolla2020,Chnng2017,
Broecker2017,Carrasquilla2017,Carleo2018,Carleo2017,Cai2018,
Choo2018,Cheng2019,Guo2018,ZYHan2018,HXie2021,Efthymiou2019}. 
The previous attempts of learning effective low-energy Hamiltonian that 
could help with accelerating the (quantum) MC sampling process, 
dubbed self-learning Monte Carlo scheme have shown the power of 
ideas from machine learning in improving the MC simulations of interacting 
fermion systems and inspired many extensions~\cite{JWLiu2017a,JWLiu2017b,
XYXu2017,Nagai2017,HLi2017,LHuang2017recommender,Endo2020,
XYXu2017PRX,ZHLiu2018,ZHLiu2019PRB,ChuangChen2018,ChuangChen2019,
ZHLiu2019PRB,ZHLiu2019PNAS,WLJiang2021}, which are, however, still
technically limited and have not enjoyed the developments of neural networks. 
On the other hand, studies of (generative) neural network models have shown 
interesting results such as normalizing flows\cite{LW2018,Hartnett2020,S-HLi2018}, 
autoregressive models\cite{DW2019,Sharir2020,JGLiu2021}, etc. and some 
networks have been successfully applied to assist MC simulations
\cite{Puente2020,Albergo2019,McNaughton2020,Singh2020,DWu2021}.

Inspired by these recent developments, we design generative neural networks 
to approximate the distribution of MC configurations, and find the network can
provide configurations with complete absence of autocorrelation, thus serving as a 
good starting point for MCMC with significantly shortened Markov chain.
We demonstrate the powerful fitting ability of the neural networks by successfully 
generating configurations from which the correct physical observables can be directly
measured after a short Markov chain, as showcased in both the classical $d=2$ 
Ising model and quantum critical point of $d=2$ fermion Hubbard model. In particular, 
the latter cannot be mapped into a classical model with short-range interactions due to 
the existence of gapless Dirac fermions, but can still be successfully dealt with by 
the network. This suggests that our approach can be applied to quantum 
many-body systems, in particular the interacting fermion models.

Moreover, we design a network-initialized Monte Carlo (NIMC) scheme with 
the assistance of such neural networks, which ``heals'' the initial bias due to the 
unthermalized training set of the neural networks and thus provides very accurate 
results. We find the NIMC scheme with independent samplings can accelerate 
simulations by reducing the long thermalization time and measurement processes,
which is especially important for simulating systems with long autocorrelation time 
in traditional MCMC.

\begin{figure}[t]
	\centering
	\includegraphics[width=9cm]{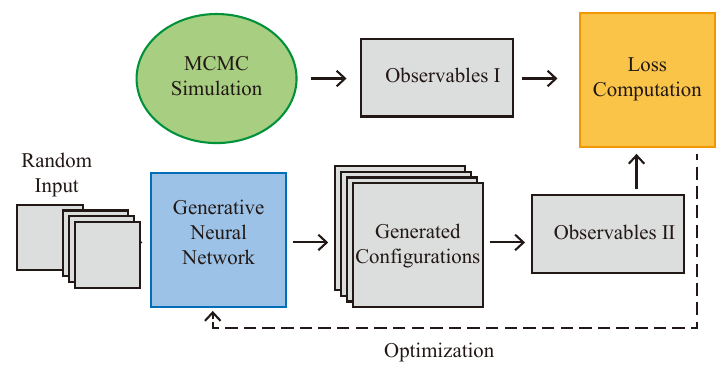}
	\caption{Flow diagram for training the generative neural networks in this paper. Observables \uppercase\expandafter{\romannumeral1} and observabless \uppercase\expandafter{\romannumeral2} refer to observables measured from training samples and from generated configurations respectively. The other terms are as explained in the main text and the loss functions are defined in Eqs.~\eqref{eq:eq2} and \eqref{eq:eq4}.}
      \label{fig:fig1}
\end{figure}

{\it Models}\,---\, In this work, we aim to train neural networks to generate MC configurations that approach the distribution of the original Hamiltonian using convolutional and transposed convolutional architectures. Different from the reference work such as autoregressive models and referring to the idea of the Generator in Generative Adversarial Nets (GAN)~\cite{Goodfellow2014}, we use random configurations (each element of the configurations is randomly set to be $\pm 1$ with equal probability) as input, which, after operations inside the networks, turn into the output of MC configurations distributed according to the physical Hamiltonian. The difference of our approach from GAN is that we do not need to build any discriminator, but directly compare the physical observables, like the internal energy and
magnetization (or magnetic structure factor) measured from a batch of configurations generated by the neural network with those obtained from MC simulations to optimize the parameters inside neural networks, as demonstrated in the flow diagram for training in Fig.\ref{fig:fig1}. 

\begin{figure}[htp!]
	\centering
	\includegraphics[width=\linewidth]{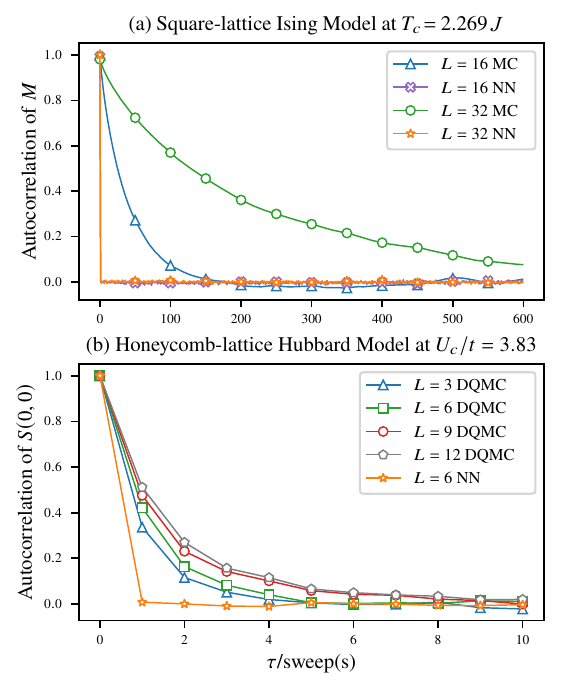}
	\caption{Comparison of autocorrelation functions (here NN represents configurations generated from neural networks and (DQ)MC refers to samplings from Markov chains). (a) shows the autocorrelation of magnetization $M$ for 2d Ising model from both MCMC and NN generated results at critical point $T_c = 2.269 J$ for $L=16, 32$. The MC results exhibit the typical critical slowing down whereas the NN results have zero autocorrelation, i.e., direct $i.i.d.$ sampling. (b) shows the autocorrelation for $S (0, 0)$ of Hubbard model on honeycomb lattice at the Gross-Neveu QCP ($U_c / t = 3.83$) from DQMC with $L = 3, 6, 9, 12$, the critical slowing down manifest. The NN results again show no sign of autocorrelation.}\label{fig:fig2}
\end{figure}

For the 2d square lattice Ising model with
\begin{equation}
H = - J \sum_{\langle i,j \rangle}\sigma_i\sigma_j\label{eq:eq1}
\end{equation}
we study the system at the critical point ($T_c\approx2.269 J$) with system sizes of $16\times 16$ and $32 \times 32$. We design the loss function directly related to the physical observables, i.e.,
\begin{equation}
\begin{aligned}
Loss&(\boldsymbol{G}_l;M_l,E_l)=w_1\sum_l[|M(\boldsymbol{G}_l)|-|M_l|]^2
\\&+w_2 \sum_l[M^2(\boldsymbol{G}_l)-M^2_l]^2+w_3 \sum_l[E(\boldsymbol{G}_l)-E_l]^2\label{eq:eq2}
\end{aligned}
\end{equation}
where $\boldsymbol{G}_l$ is the $l$-th generated configuration, 
$M_l$ and $E_l$ refer to the magnetization $\frac{1}{N}|\sum_j \sigma_{j}|$ and 
the energy density $\frac{1}{N}\langle H\rangle$ measured from the corresponding 
$l$-th MCMC configuration,
and $w_1$, $w_2$, and $w_3$ are constants to balance each part of the loss to be in the same magnitude.

For the 2d Hubbard model with
\begin{equation}
H=-t\sum_{\langle i,j\rangle,\sigma}(c_{i,\sigma}^{\dagger}c^{\phantom{\dagger}}_{j,\sigma}+\mathrm{h.c.})+U\sum_i(n_{i,\uparrow}-\frac{1}{2})(n_{i,\downarrow}-\frac{1}{2}) \label{eq:eq3}
\end{equation}
 we consider the half-filled honeycomb lattices where it experiences a chiral Heisenberg Gross-Neveu quantum critical point at $U_c/t \approx 3.83$ between the Dirac semimetal and the antiferromagnetic Mott insulator~\cite{Meng2010,Sorella2012,Assaad2013,TCLang2019,YZLiu2020}. Here the configuration is made of the auxiliary fields used in the
discrete Hubbard-Stratonovich transformation to decouple the fermion interaction terms in DQMC, 
so the configurational space is of the size $L\times L\times \beta$ compared with the $L\times L$ 
for the Ising case (see detailed explaination in Sec. I of Supplemental Materials (SM)~\cite{suppl}). 
We then design the loss function as:
\begin{equation}
\begin{aligned}
Loss(\boldsymbol{G}_l; S(\boldsymbol{Q})_l,E_{k_l})&=w_1\sum_l[S(\boldsymbol{Q})(\boldsymbol{G}_l)-S(\boldsymbol{Q})_l]^2
\\&+w_2\sum_l[E_k(\boldsymbol{G}_l)-E_{k_l}]^2,
\label{eq:eq4}
\end{aligned}
\end{equation}
where $\boldsymbol{G}_l$ is the $l$-th generated configuration, $S(\boldsymbol{Q})=\frac{1}{N} \sum_{ij}e^{-i\mathbf{Q} \cdot (\mathbf{r}_i-\mathbf{r}_j)}\langle s_i^z s_j^z\rangle$ ($s^{z}_{i}=\frac{1}{2}(n_{i,\uparrow}-n_{i,\downarrow})$ refers to the magnetic structure factor (so $S(\boldsymbol{Q})_l$ and $S(\boldsymbol{Q})(\boldsymbol{G}_l)$ refer to the structure factor measured from the $l$-th training sample and from the $l$-th generated configuration)  and $S(\Gamma=(0,0))$ is chosen for honeycomb lattice as the $\Gamma=(0,0)$ is the ordered wave vector for the antiferromagnetic long-range order, and $E_k$ is the kinetic energy $\frac{1}{N} \sum_{\langle i,j \rangle,\sigma}\langle( c^{\dagger}_{i,\sigma} c_{j,\sigma}+ h.c.)\rangle$, with constants $w_1$ and $w_2$.

The design of the loss functions comes from physical intuition, and we choose the key physical observables such as the energy and magnetic structure factor, which reflect the phase transition of our interests in the loss function. Similar considerations can be generalized to other problems. In practice,
our networks are all based on Tensorflow~\cite{Abadi2016}, and we complement with 
more details of the Ising and Hubbard models in Sec.I and those of neural networks in 
Sec.II of SM~\cite{suppl}.

{\it Autocorrelation analysis}\,---\,
In MCMC, the autocorrelation time, associated with the particular update
scheme, can be extremely long in the physically interesting parameter regime
like, e.g., near the classical and quantum critical points. And it is even worse 
when one performs the finite size analysis as the autocorrelation time usually 
increase with system size with a high power~\cite{SwendsenWang1987,XYXu2017,ChuangChen2018}. 
It therefore benefits a lot that the independent and direct samplings from 
the neural networks are completely free of autocorrelation. We demonstrate these 
behaviours quantitatively by measuring the autocorrelation function 
(defined in Sec.I.3 of SM~\cite{suppl}) of magnetization for 2d Ising model at  
$T_c=2.269 J$ in Fig.~\ref{fig:fig2}(a), and of $S(0,0)$ for the honeycomb-lattice Hubbard model
at the Gross-Neveu QCP ($U_c/t=3.83$) in Fig.~\ref{fig:fig2}(b), 
between the results generated by neural networks and MCMC results. 
It can be observed that, for the MCMC the autocorrelation functions decay slowly 
in Monte Carlo steps and the autocorrelation time for classical (quantum) critical points 
increases with the system size, but in results from neural networks, 
such autocorrelation has been completely eliminated.

{\it NIMC algorithm based on neural networks}\,---\,
We now introduce the NIMC algorithm.
In order to demonstrate its ability of correcting the error in the training sample, we no longer train the networks to approximate
well-thermalized MC results, but take unthermalized MC configurations generated from short Markov chains. After training the neural
networks and generate independent configurations, we add MCMC updates starting from these generated configurations (one Markov chain for each configuration)~\footnote{For direct measurement on the generated configurations, we map each generated number from the range of $(0,1)$ to $(-1,1)$ and the relationship is: ${\rm number\ for\ measurement} = {\rm generated\ number}*2-1$. For MCMC update, we use Bernoulli distribution to discretize each element with the initial generated number as the possibility to take the value $1$, and map every $0$ to $-1$. } and find the NIMC results quickly converge to the well-thermalized results.
The whole process is summarized in Alg.~\ref{alg:alg1}.

\begin{algorithm}[H]
1: Run a short Markov-chain (unthermalized) as training samples\\
2: Train a generative neural network using observables from training samples\\
3: Generate configurations from the trained neural network\\
4: Start short MCMC steps from the generated $i.i.d.$ configurations and make measurements
\caption{NIMC scheme based on neural networks}
\label{alg:alg1}
\end{algorithm}

\begin{figure}[t]
\centering
\includegraphics[width=\linewidth]{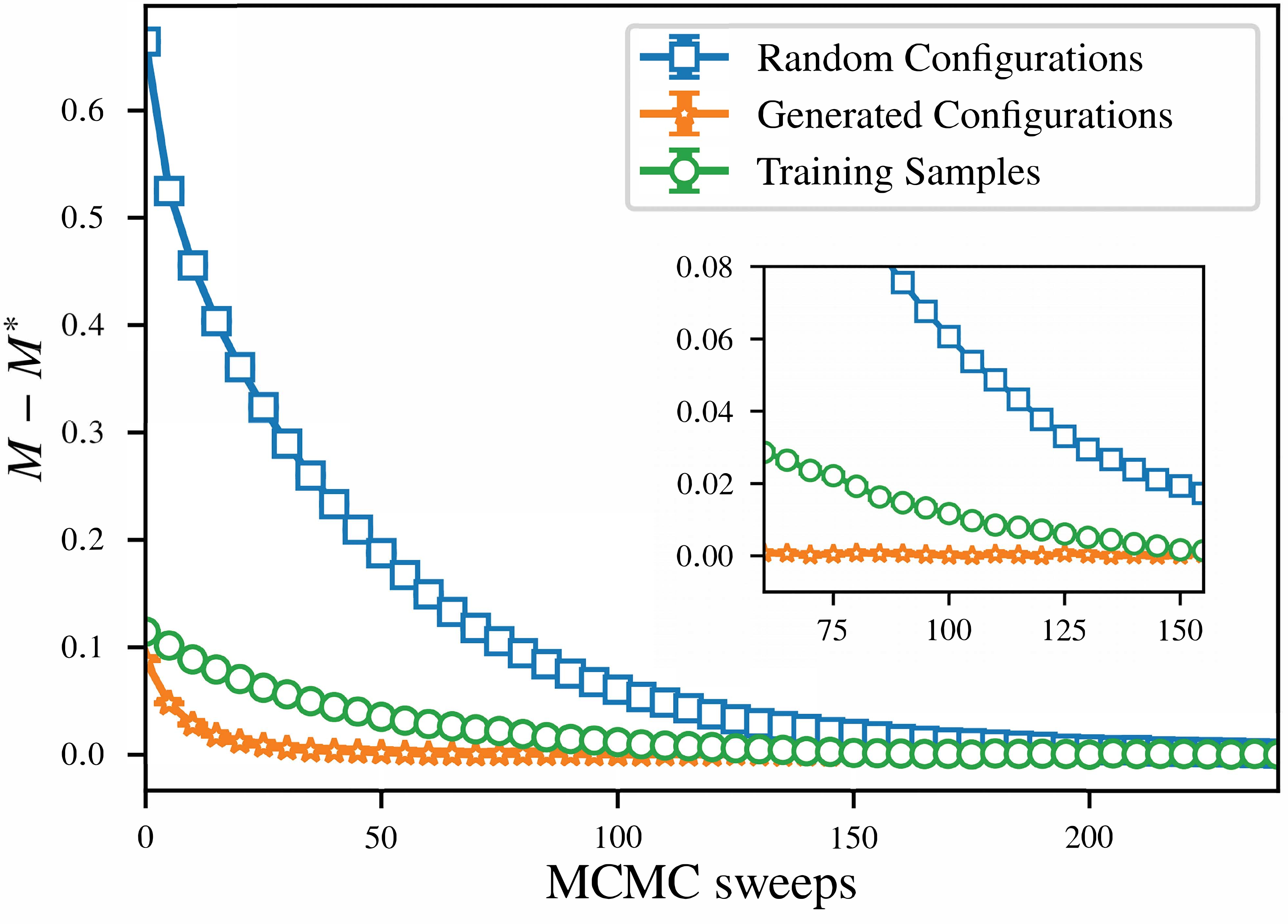}
\caption{Comparing the convergence time of magnetization
	(MCMC sweeps needed to converge) between NIMC
	and the two MCMC simulations of random
	configurations and configurations used for traning the neural network,
	respectively. We set $M^*=0.713$ for reference. For all three sets
	(each consisted of 1000 configurations), we run 50 iterations from
	the beginning to derive the expectations with standard errors.
	The inset further highlight the superiority of NIMC and the scale
	on $y$-axis is not changed. }\label{fig:fig3}
\end{figure}

\begin{table}[h]
\caption{ Numerical results of NIMC on square lattice Ising model ($L=16$, $T_c=2.269 J$). The four sets of data respectively refer to training samples (unthermalized), neural network generated results, NIMC results (30 more MCMC sweeps), and well-thermalized MC results. Results shown here are expectations with standard errors from 1000 values (and good MC takes 1000 bins).}\label{tab:tab1}
\centering
\begin{tabular}{p{16mm}<{\centering}p{16mm}<{\centering}p{16mm}<{\centering}p{16mm}<{\centering}p{16mm}<{\centering}}
\toprule
Metric & MC Train & NN & NIMC & Good MC \\
\hline
$|M|$ & 0.600(7) & 0.606(3) & 0.713(6) & 0.713(3)\\
$E$    & -1.373(6) &  -1.329(4) & -1.454(5) & -1.453(2)\\
\bottomrule
\end{tabular}
\end{table}

{\it Numerical experiments on the Ising model}\,---\,
In the standard MCMC, the Markov chains must first be well thermalized such that it converges to the correct distribution, and then generate enough bins of statistically uncorrelated configurations to yield expectation values of observables with
satisfactory statistical error, which requires the thermalization and each bin to be longer than the system-size-dependent autocorrelation time. To demonstrate 
that NIMC largely reduces the thermalization process, we compare the 
performance of NIMC on $16 \times 16$ Ising model at critical point, with those starting MCMC simulations from random configurations and configurations used as training samples in NIMC, by computing the MCMC sweeps (local update) needed for converged  magnetization results ($M^*=0.713$ for $L=16$ and $T=T_c$ as reference). We run 50 iterations for all three sets of data with each set consisted of 1000 configurations (for each iteration, we re-run Markov chains from all the configurations), and the
results are shown in Fig.~\ref{fig:fig3}. For training samples, almost
about 150 sweeps are needed to reach a convergence,
and for random configurations the number is more than 200. Remarkably, the NIMC scheme greatly
lowers the thermalization time and arrive at the convergence after about only 30 sweeps.

To further quantify the comparison, we take the training part for neural networks into consideration, and compute on the same CPU (Intel Xeon E5-2680) that for the 150 epochs of network training that are more than sufficient to converge the parameters in the neural network. It turns out that the training process takes around 125 seconds (could be made even faster on GPU), and each MCMC sweep for 1000 configurations takes around 2.5 seconds. That is, NIMC takes
$125+2.5\times30=200$ seconds for the 1000 thermalized and independent configurations here from training samples, while the $1000$ samples themselves need up to $2.5\times150=375$ seconds for thermalization. 
Therefore, we claim that NIMC can overall speed up the MC simulations. Besides, the independent NIMC samples also save time in the process of measurements as there is
completely no autocorrelation and we can directly measure observables
once the thermalization is completed.
%

\begin{figure}[t!]
	\centering
	\includegraphics[width=\linewidth]{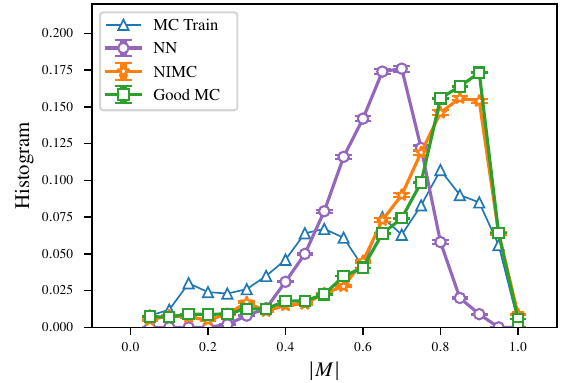}
	\caption{Histograms of the distribution of magnetization and the abbreviations are the same as in Table.\ref{tab:tab1}. We obtain standard errors from 50 iterations of NIMC (30 more MCMC sweeps and 1000 configurations each) in total with MC Train fixed, and the Good MC are from 50 sets (each set consists of 200000 values).}\label{fig:fig4}
\end{figure}

To better compare the accuracy, we then show the 
 observables by taking 1000 configurations from the beginning of a 
rather short Markov-chain as traing samples to run NIMC and the numerical results 
are shown in Tab.~\ref{tab:tab1}. Besides, we also plot the corresponding histograms 
of distribution of magnetization in Fig.~\ref{fig:fig4} after 50 iterations of NIMC 
(1000 configurations for each iteration) from the same training samples. 
The observables obtained from unthermalized training samples are far 
worse than those thermalized. After training, the neural network generates 
configurations with approximately the same observables but of Gaussian-like 
distributions. Here, we think this change of distribution is the reason for the faster convergence for MCMC simulations. After we add 30 MCMC updates on the generated configurations, 
they quickly converge to the results of the so-called Good MC, which consists of totally 
200000 well thermalized MC configurations for reference (then we take 1000 bins with 200 
configurations per bin). Moreover, the distribution of NIMC has a large overlap with Good 
MC. To quantitatively analyze the overlap of distribution, we compute the percentage overlap 
($\%OL$) between two histograms~\cite{Singh2020}, i.e., $\%OL (P_r,P_{\theta})=\sum_i 
min(P_r(i),P_{\theta}(i))$, where $P_r$ and $P_{\theta}$ corresponds to the distributions to 
be compared and we divide them into 20 bins. Again we count the NIMC samples from 50 
iterations for statistical expectation with standard error. The result is $\%OL=94.78(18)$, 
which reconfirms that the distribution of NIMC with just 1000 configurations has a large 
overlap with that of Good MC.

\begin{figure}[t]
	\centering
	\includegraphics[width=0.89\linewidth]{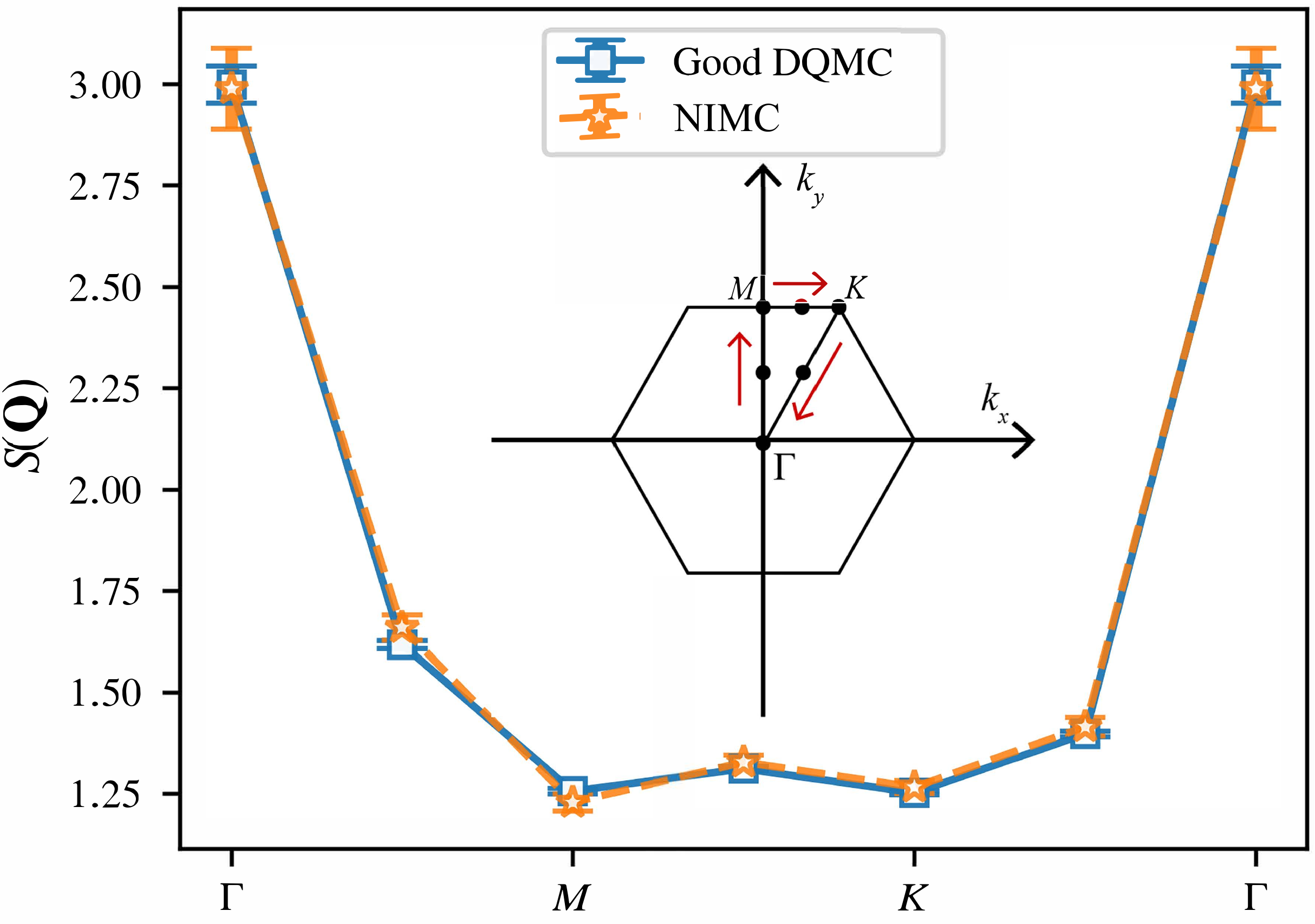}
	\caption{Comparison of magnetic structure factor $S(\boldsymbol{Q})$ between Good DQMC (10000 samples) and NIMC (15 sweeps), which are not used for network training except $S(\Gamma)$. The black dots in the inset are the momenta measured along the high-symmetry path (with red arrows).}\label{fig:fig5}
\end{figure}

\begin{table}[h]
\caption{ Numerical results of NIMC on honeycomb lattice Hubbard model at 
$(U_c/t=3.83, L=6)$. The three sets of data respectively refers to training samples (unthermalized), NIMC results, and well-thermalized DQMC results. Results 
shown here are expectations with standard errors from 1000 values(Good DQMC
take 1000 bins).}\label{tab:tab2}
\centering
\begin{tabular} {p{20mm}<{\centering}p{20mm}<{\centering}p{20mm}<{\centering}p{20mm}<{\centering}}
\toprule
Metric & DQMC Train &  NIMC & Good DQMC \\
\hline
$E_k$ & -1.360(4) &  -1.357(3) & -1.358(1)\\
$S(\Gamma)$    & 3.11(10) &   2.99(10) & 3.00(4)\\
\bottomrule
\end{tabular}
\end{table}

{\it Hubbard model on the honeycomb lattice}\,---\,
Beyond the classical Ising model, here we take the 2d Hubbard model on a 
honeycomb lattice at its unique Gross-Neveu quantum criticality between the 
Dirac semimetal and antiferromagnetic Mott insulator $(U_c/t=3.83, L=6)$ as 
an example of interacting fermion systems. We run NIMC with 15 DQMC sweeps 
from 1000 generated configurations and the mean values with errorbars of physical 
observables are listed in Tab.~\ref{tab:tab2}. The NIMC results again present 
almost identical expectation values of physical observables as the Good DQMC 
(containing 10000 configurations in total that are divided 
into 1000 bins). Furthermore, in Fig.~\ref{fig:fig5} we show the magnetic structure 
$S(\mathbf{Q})$ along the high-symmetry path of the Brilliouin zone, which are 
observables not used in training the networks. We find again excellent 
agreements between Good MC and NIMC only after 15 sweeps. 
The peak at the $\Gamma$ point highlight the quantum critical fluctuations 
towards the magnetic order that will gap out the Dirac cones. These results 
confirm the neural networks with optimizers like ADAM~\cite{Kingma2015} are 
capable of fixing the large-scale determinant computations. We have to clarify 
that in NIMC we do not necessarily need up to 1000 
configurations (as we use for example here), which can further reduce the 
computational cost.

{\it Discussion}\,---\, By developing a NIMC method based on the generative neural networks, our numerical results support firmly that the latter as independent and direct sampling approach can approximately capture the weight distributions of MC configurations and completely get rid of autocorrelation in the study of classical and quantum many-body systems. NIMC therefore provides a scheme for speeding up the MC simulations in reducing the thermalization time and saving time for direct measurement of observables from those independent NIMC samples. Admittedly, our NIMC still needs a few steps of the MCMC simulation, 
therefore it does not change the overall scaling of the computational cost
but can nevertheless lead to a significant factor reduction.

From the perspective of neural networks, we verify that large-scale simulations, such as DQMC calculations, can be implemented through the evaluation
of the neural network loss functions.
The next step from here is to use generative neural networks to provide efficient 
and accurate system-size extrapolations of MC configurations and in this way many 
more complicated yet fundamental problems, such as the quantum moir\'e materials that the DQMC has just been shown to be able to solve 
~\cite{XuZhang2021,Liao2021CPB,JYLee2021,Liao2021PRX,
Liao2019PRL,GPPan2021,XuZhang2021SC} but with heavy computational costs, 
could be improved in NIMC.

{\it Acknowledgments}\,---\,
We thank Zheng Yan, Shangqiang Ning, Bin-bin Mao, Jiarui Zhao, Chengkang 
Zhou, and Xu Zhang for the enjoyable discussions and happy conversations 
in the No. 16 Pavillion of Lung Fu Shan Country Park. We acknowledge
support from the RGC of Hong Kong SAR of China (Grant Nos. 17303019,
17301420, 17301721 and AoE/P-701/20), NSFC (Grant Nos.~11974036, 11874115 and
11834014), the Strategic Priority Research Program of the Chinese Academy 
of Sciences (Grant No. XDB33000000) and the K. C. Wong Education Foundation (Grant No. GJTD-2020-01). We thank the Center for Quantum
Simulation Sciences in the Institute of Physics, Chinese Academy of Sciences, 
the Computational Initiative at the Faculty of Science and the Information 
Technology Services at the University of Hong Kong for their technical support 
and generous allocation of GPU/CPU time. This work is also supported 
by the Seed Funding "Quantum-Inspired explainable-AI" at the HKU-TCL Joint
Research Centre for Artificial Intelligence, Hong Kong.

\begin{widetext}

	\section{Physical Models}
	
	\subsection{1.   \ \ Model Details}
	First we consider two-dimensional Ising model on square lattice at the critical point $T_c\approx2.269 J$  and the Hamiltonian is shown in Eq.(1) of the main text, which can be solved with classica Monte Carlo simulations. $\sigma_i$ and $\sigma_j$ refers to nearest neighbors of classical spins and take values: $\pm1$. We set the interaction strength $J$ of all pairs of neighbors to be $1$.
	
	Next, we consider Hubbard model on the honeycomb lattice, which can be solved with determinant quantum Monte Carlo (DQMC) at half-filling.  In Eq.(3) of main text, $t$ is the hopping parameter for the kinetic energy, $U$ is the repulsive Coulomb interaction between electrons on the same lattice site, $\langle i,j \rangle$ represents a pair of nearest-neighbor sites in the lattice, the operators $c_{i,\sigma}^{\dagger}$ and $c_{i,\sigma}$ are the fermion creation and annihilation operators for fermions with z component of spin-up ($\sigma=$ $\uparrow$) or spin-down ($\sigma=$ $\downarrow$), and the operators $n_{i,\sigma}=c_{i,\sigma}^{\dagger}c_{i,\sigma}$ are the number operators which count numbers of fermions of spin $\sigma$ on the site $i$.  In loss function Eq.(4) in the main text, $S$ refers to the structure factor:
	$S(\boldsymbol{Q}) =\frac{1}{L^2} \sum_{ij}e^{-i\mathbf{Q} \cdot (\mathbf{r}_i-\mathbf{r}_j)}\langle s_i^z s_j^z\rangle$, and the spin operator $\vec{s}_i=\frac{1}{2}\sum_{\alpha\alpha^{'}}c_{i,\alpha}^\dagger \overrightarrow{\sigma}_{\alpha\alpha^{'}} c_{i,\alpha^{'}}$. Note here $\overrightarrow{\sigma}$ denotes the Pauli matrices and so the z-component $s_i^z=\frac{1}{2}(n_{i\uparrow}-n_{i\downarrow})$.

	\subsection{2.   \ \ Determinant Quantum Monte Carlo (DQMC)}
	The partition function $Z = \Tr \{e^{-\beta H}\}$ is expressed as a path integral by discretizing the inverse temperature $\beta$ into $L_{\tau}$ slices of length $\Delta\tau$. Then, after Trotter-Suzuki decomposition, the Hamiltonian are separated in each time slice and $Z$ can be written as 
	\begin{equation}
		Z= \Tr \left[ \prod_{l=1}^{L_{\tau}} e^{-\Delta\tau H_k}e^{-\Delta\tau H_U} \right] + O(\Delta\tau^2),
	\end{equation}
	where $H_k=-t\sum_{\langle i,j\rangle,\sigma}(c_{i,\sigma}^{\dagger}c_{j,\sigma}+h.c.)$ is the kinetic term and $H_U=U\sum_i(n_{i,\uparrow}-\frac{1}{2})(n_{i,\downarrow}-\frac{1}{2})$ is interaction term of the Hubbard model. 
	
	To treat the quartic interaction term, the discrete Hubbard-Stratonovich transformation[1-3] can be applied and then $e^{-\Delta\tau H_U} = e^{-U\Delta\tau (n_{i,\uparrow}-\frac{1}{2})(n_{i,\downarrow}-\frac{1}{2})} = \frac{1}{2}e^{-\frac{U\Delta\tau}{4}} \sum_{s_i=\pm1}e^{\nu s_i(n_{i,\uparrow}-n_{i,\downarrow})}$, where the scalar $\nu$ is defined by $\cosh{\nu}=e^{\frac{U\Delta\tau}{2}}$.
	Putting kinetic part and interaction part together, the term in partition function becomes
	\begin{equation}
		e^{-\varDelta\tau H_k} e^{-\varDelta\tau H_U} = \prod_{\sigma} e^{-\varDelta\tau T_\sigma} e^{\sigma \nu s_i n_{i,\sigma}},
	\end{equation}
	where $H_k$ is rewritten with operator $T_\sigma = -t\sum_{i,j}c_{i,\sigma}^\dagger c_{j, \sigma} + h.c.$. 
	According to the feature of fermion operator[1], a fermion operator ($\hat{M}_l$ for example) with a quadratic form like $\hat{M}_l = \sum_{i,j} c_i^\dagger(M_{l})_{ij}c_j$ satisfies
	\begin{equation}
		\Tr\left[ e^{-\hat{M}_1} e^{-\hat{M}_2} \cdots e^{-\hat{M}_L} \right] = \Det[I+e^{-M_1 }e^{-M_2 }\cdots e^{-M_L}].
	\end{equation}
	Then the partition function could be finally written in determinant form as
	\begin{equation}
		Z = \left(\frac{1}{2}e^{-\frac{U\Delta\tau}{4}} \right)^{NL_{\tau}} \sum_{s_{i,l}} \prod_{\sigma} \Det\left[ I + B^{\sigma}(L_\tau, L_{\tau} -1)\cdots B^{\sigma}(1, 0) \right],
	\end{equation}
	in which 
	\begin{equation}
		B^{\sigma}(l_2, l_1) = \prod_{l=l_1+1}^{l_2} e^{\sigma \nu \Diag\{s_{i,l}\}} e^{-\varDelta\tau T},
	\end{equation}
	where $\sigma = \{1, -1\}$ in calculation corresponding mark $\{\uparrow, \downarrow\}$, and $T$ is the matrix corresponding to the operator $T_\sigma$.
	Thus, we have introduced a sum over the field of auxiliary variables $s_{i,l}$ in a $(d+1)$-dimension space ($d$ for spatial denoted by $i$ and 1 for imaginary time denoted by $l$) as shown in Fig.~\ref{fig:fig6} (b), and the fermionic degrees of freedom in the quadratic form have been integrated out analytically. Note $B^{\sigma}$ is an $N \times N$ matrix that depends on the auxiliary configurations.
	
	\subsection{3.   \ \ Autocorrelation Function}
	The autocorrelation function for an observable $O$ is defined as:
	\begin{equation}
		A_O(t)=\frac{\langle O(i+t)O(i) \rangle-\langle O\rangle^2}{\langle O^2\rangle-\langle O\rangle^2}.
		\label{eq:eq5}
	\end{equation}
	For MCMC, $i$ and $t$ denotes the simulation time, normally in units of the MC sweeps (one sweep means doing flipping attempts over all the spins of the configuration), and the averages are over the reference time $i$. For Neural Network results, when computing the formal autocorrelation function, $i$ and $t$ simply refers to the serial numbers of the configurations.
	
	\section{Neural Network Details}
	
	\subsection{1.   \ \ Architecture}
	
	\begin{figure*}[htp!]
		\centering
		\includegraphics[width=\textwidth]{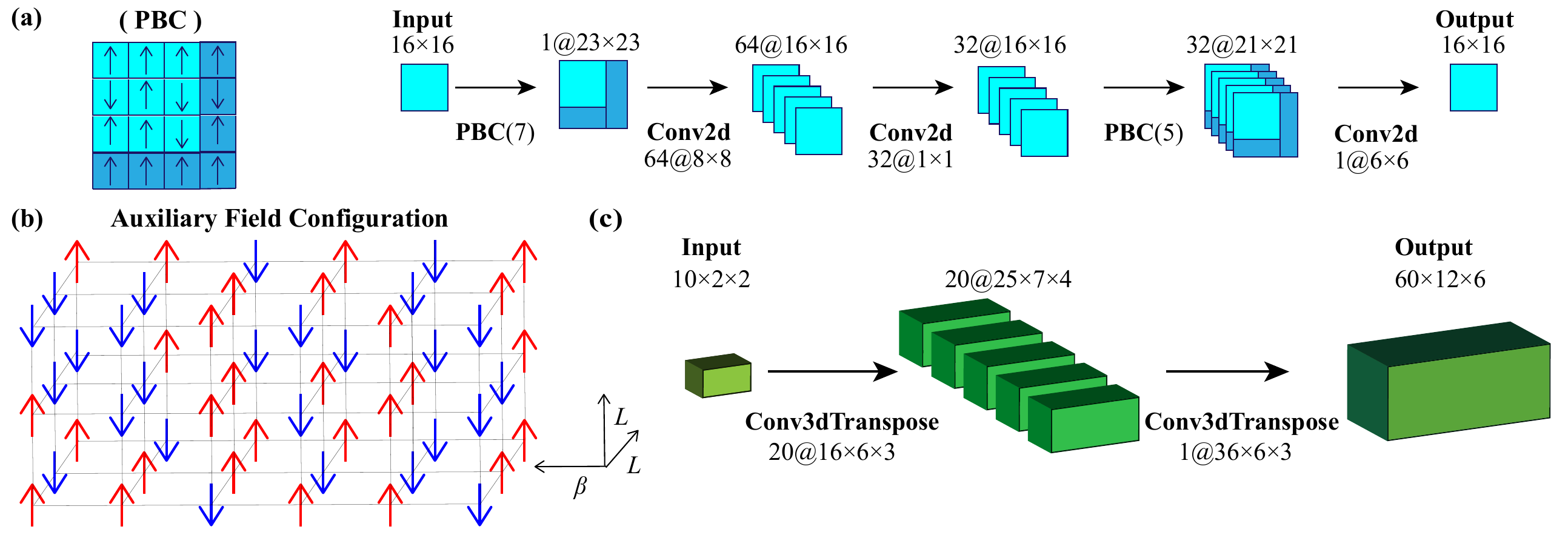}
		\caption{Schematic figures for the generative neural networks used in this study. (a) shows an example of periodic boundary condition on the left, and we use the neural network structure on the right for the training of the 2d Ising model, where Conv2d stands for 2d convolutional layer. (b) displays the auxiliary field configuration in DQMC for the Hubbard model, with the $(2+1)d$ space-time of $\beta\times L\times L$. The auxiliary fields of $\pm 1$ live on each space-time lattice site. (c) demonstrates the schematic structure of 3d transposed convolutional network we employed for the training of the 2d Hubbard model on $6\times6 \times2$ honeycomb lattice.}
		\label{fig:fig6}
	\end{figure*}
	
	For 2d Ising model on square lattice ($16\times 16$ for example), the network structure is shown in Fig.~\ref{fig:fig6} (a). For the random input, we fix its shape to be the same as that of the Monte Carlo configurations. Inside the network, inspired by the network structure in Ref.[30], we use three 2d convolutional layers, with 64 filters of size $8 \times 8$ for the first layer, 32 filters of size 1 for the second, and 1 filter of size 6 $\times$ 6 for the third layer. We choose rectified linear function ReLU(x) = max(0, x) in the first and second convolutional layers and sigmoid function $\sigma(x) = 1/(1+e^{-x})$ in the third. We also apply periodic boundary condition (PBC) layers each time before the convolutional layers if the kernel dimension is larger than one, which provides the configuration tensors with paddings of boundary elements instead of 0s (the padding size is decided by the size of filters in the corresponding convolutional layer, so as to ensure the shape invariance of the configuration tensors after convolutional operations).
	
	In the case of 2d Hubbard model on the honeycomb lattice ($6\times 6\times 2$ for example), we switch from classical Ising configurations to the auxiliary field configurations of the DQMC[3,33], with the configuration space of $\beta\times L\times L \times 2$ where $\beta=1/T$ is the inverse temperature. We set $\beta=6$ and the Trotter discretization $\Delta \tau=0.1$, so the DQMC auxiliary field configurations are of shape 60 $\times$ 12 $\times$ 6 (note the 2 site per unit cell for the honeycomb lattice), as schematically shown in Fig.~\ref{fig:fig6} (b). As shown in Eq.(4), we have to carry out large-scale determinant computation to measure the observables from generated configurations in order to compute the loss, and thus the complexity for the optimization is much larger than the case of classical model. Therefore, instead of the previous PBC-based network structure, we try another architecture of transposed convolutional layers and feed smaller random configurations as input. We find it is still capable of fitting the results we want and optimize faster. As shown in Fig.~\ref{fig:fig6} (c) for the example model, we build two 3d transposed convolutional layers, with all filters using valid padding and sigmoid function as activation. There are 20 filters of size 16 $\times$ 6 $\times$ 3 in the first layer and 1 filter of size 36 $\times$ 6 $\times$ 3 in the second and the random input configurations are of shape 10 $\times$ 2 $\times$ 2 while the predicted output are of shape  60 $ \times$ 12 $\times $ 6.

	\subsection{2.   \ \ Optimization}
	In order to optimize such neural networks, for both classical and quantum models, we prepare 1000 sets of observables measured from MC simulations and take 1000 input random configurations which will be processed into generated configurations. The comparison in loss function is randomly distributed without any grouping. For the choice of observables in the defined loss functions, we simply pick some from the ones we usually focus on. In the Ising case, we take the batch size to be 5 and epoch number to be up to 150 as the computation is quite easy. While in the Hubbard case, we run at most 15 epochs with a batch size of 3. We optimize the network parameters using Adam[63] with conventional learning rate $10^{-3}$,  $\beta_1=0.9$, and $\beta_2=0.999$.
	
\end{widetext}

\setcounter{equation}{0}
\setcounter{figure}{0}
\renewcommand{\theequation}{S\arabic{equation}}
\renewcommand{\thefigure}{S\arabic{figure}}

\end{document}